\documentclass[conference]{IEEEtran}
\IEEEoverridecommandlockouts

\usepackage{cite}
\usepackage{amsmath,amssymb,amsfonts}
\usepackage{algorithm}
\usepackage{algpseudocode}
\usepackage{graphicx}
\usepackage{textcomp}
\usepackage{xcolor}
\usepackage{tabularx}
\usepackage{physics}
\usepackage{hyperref}
\usepackage[a4paper, total={184mm,239mm}]{geometry}

\def\BibTeX{{\rm B\kern-.05em{\sc i\kern-.025em b}\kern-.08em
    T\kern-.1667em\lower.7ex\hbox{E}\kern-.125emX}}

\begin{document}
\bstctlcite{IEEEexample:BSTcontrol}

\title{
Quantum Circuit Simulation with \\Fast Tensor Decision Diagram
}

\author{
\IEEEauthorblockN{Qirui Zhang, Mehdi Saligane, Hun-Seok Kim, David Blaauw, Georgios Tzimpragos, Dennis Sylvester}
\IEEEauthorblockA{
Department of Electrical Engineering and Computer Science, University of Michigan, Ann Arbor, MI, USA\\
\{qiruizh, mehdi, hunseok, blaauw, gtzimpra, dmcs\}@umich.edu
}
}

\maketitle

\begin{abstract}

Quantum circuit simulation is a challenging computational problem crucial for quantum computing research and development. The predominant approaches in this area center on tensor networks, prized for their better concurrency and less computation than methods using full quantum vectors and matrices. However, even with the advantages, array-based tensors can have significant redundancy. We present a novel open-source framework that harnesses tensor decision diagrams to eliminate overheads and achieve significant speedups over prior approaches. On average, it delivers a speedup of 37$\times$ over Google's TensorNetwork library on redundancy-rich circuits, and 25$\times$ and 144$\times$ over quantum multi-valued decision diagram and prior tensor decision diagram implementation, respectively, on Google random quantum circuits. To achieve this, we introduce a new linear-complexity rank simplification algorithm, Tetris, and edge-centric data structures for recursive tensor decision diagram operations. Additionally, we explore the efficacy of tensor network contraction ordering and optimizations from binary decision diagrams.

\end{abstract}


\section{Introduction}

Quantum circuit simulation (QCS) is essential for various aspects of quantum computing research and development, including software development and debugging~\cite{b8}, noise simulation~\cite{b9, b10}, quantum chip validation~\cite{b11}, and hybrid quantum-classical algorithms~\cite{b12}. Meanwhile, QCS has inherent scalability limitations due to its exponential complexity, particularly when dealing with deep and random circuits~\cite{b11}. To overcome the obstacle, researchers have been actively seeking solutions to reduce the otherwise unacceptably long runtime. Among the solutions are $\text{Schr\"odinger}$-style methods that utilize full vector and matrix representations~\cite{b8, b13, b16} as well as tensor network (TN) approaches~\cite{b17, b18, b19}, with the latter demonstrating a clear advantage via a range of optimizations to improve concurrency and reduce complexity, such as TN partitioning and index slicing~\cite{b19}, tensor decomposition~\cite{b20}, hyper-edges~\cite{b19}, and contraction ordering~\cite{b21}.

While significant progress has been made in recent years, matrix and tensor still contain a considerable amount of redundancy when using arrays~\cite{b25, b26}. In response to this, a small but growing number of studies have been looking at decision diagrams as a promising alternative to arrays. For example, quantum information decision diagram (QuIDD)~\cite{b22} and quantum multi-valued decision diagram (QMDD)~\cite{b23, b24, b25} compress arrays by encoding them as directed acyclic graphs (DAGs). This indeed reduces redundancy, which is the primary objective. However, QuIDD and QMDD are specifically developed for $\text{Schr\"odinger}$-style methods, and cannot leverage the numerous TN optimizations that are readily available.

More recently, tensor decision diagram (TDD) has emerged as a solution that effectively merges the strengths of both TN and DD approaches. TDD builds upon TN formalism, so TN optimizations are still applicable, and extends the idea of DD to compress tensor arrays. Despite its theoretical promise, existing implementation of TDD~\cite{b26} (referred to as PyTDD for the rest of the paper) has not yet provided compelling results to demonstrate their practical superiority. The reason for this performance gap can be attributed to the limited exploration of the potential benefits from complexity-reducing TN optimizations in this specific context, as well as the software's inefficient implementation in Python.

\begin{figure}[ht]
\centering
\includegraphics[width=\linewidth]{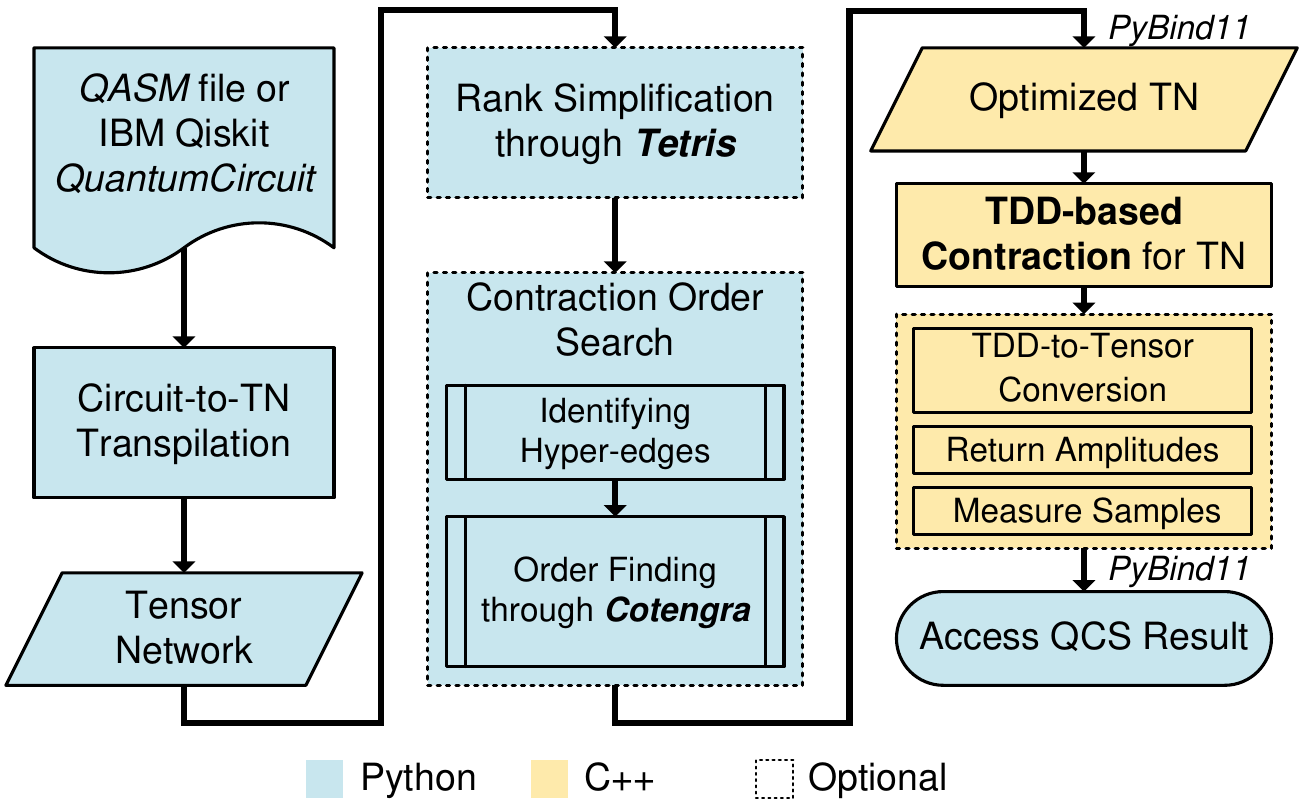}
\caption{Overview of the presented FTDD framework.}
\label{fig8}
\end{figure}

To fully harness the capabilities of TDD, we present fast tensor decision diagram (FTDD). FTDD is an open-source QCS framework that is primarily developed in C++ and incorporates a range of optimizations, drawing from both the novel techniques introduced in this paper and the extensive TN and DD literature. Fig.~\ref{fig8} provides an overview, distinguishing Python in blue and C++ in yellow. FTDD comprises two steps: the first step involves TN optimizations as preprocessing, and the second step focuses on TDD optimizations. For TN optimizations, it begins by constructing a TN, then applies a novel linear-complexity algorithm called Tetris for rank simplification, followed by an exploration of near-optimal TN contraction orders. In the second step, the information extracted from the previous step is utilized to execute near-optimal contraction orders for the rank-simpilifed TN using TDDs. To optimize TDD, we propose edge-centric data structures aimed at eliminating overhead in recursive TDD operations and incorporate key optimizations borrowed from binary decision diagram (BDD)~\cite{b27, b30} for further efficiency.

To test our hypotheses, we compare FTDD to PyTDD, QMDD, and Google's TensorNetwork library (GTN) on various quantum circuits. On Google random quantum circuits (RQCs)~\cite{b29}, FTDD achieves a 144$\times$ speedup over PyTDD and a 25$\times$ speedup over QMDD. This makes FTDD the fastest DD-based QCS framework to date. On a collection of redundancy-rich circuits from MQT Bench~\cite{b34}, FTDD achieved an average speedup of 37$\times$ over GTN. FTDD software and test cases are publicly available at \href{https://github.com/QiruiZhang/FTDD}{https://github.com/QiruiZhang/FTDD}.

\section{Background}

This section provides an overview of the fundamentals of TN and TDD. Additionally, we introduce an example quantum circuit, Fig.~\ref{fig2}, borrowed from a Google RQC instance~\cite{b29}. This circuit, or parts of it, will serve as a recurring illustration throughout the rest of the paper. For quantum computing basics, we recommend consulting~\cite{b0}.

\subsection{Quantum Circuits as Tensor Networks}

\begin{figure}[htbp]
\centering
\includegraphics[width=\linewidth]{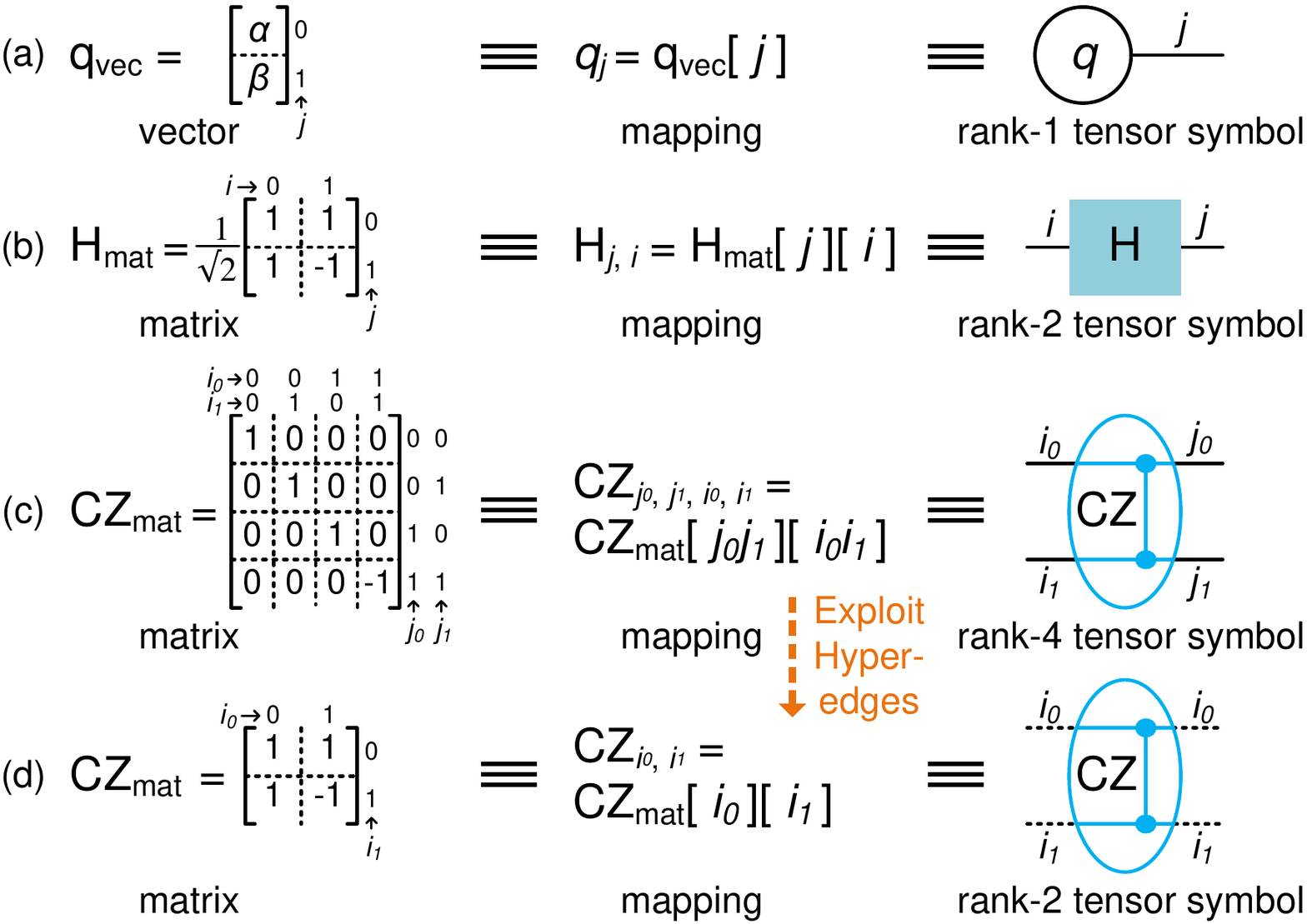}
\caption{Qubit and quantum gates represented in matrices and tensor symbols: (a) single-qubit; (b) $H$ gate; (c) $CZ$ gate; (d) $CZ$ using hyper-edges.}
\label{fig1}
\end{figure}

A tensor is a multi-dimensional array of complex numbers~\cite{b20}. In visual representation, a tensor takes the form of a shape with edges, where each edge represents a dimension identified by an index. Each index can take one of $D$ different values (0, 1, ..., or $D-1$). When simulating a qubit, these indices adopt two distinct values, 0 or 1, as shown in Fig.~\ref{fig1} (a). The rank of a tensor corresponds to the number of different dimensions it possesses.
Fig.~\ref{fig1} (b) shows how the Hadamard gate ($H$) matrix is mapped to a rank-2 tensor, and Fig.~\ref{fig1} (c) shows how the controlled-Z gate ($CZ$) matrix is mapped to a rank-4 tensor. For $CZ$, $i_{0}$ and $j_{0}$ are respectively input and output for the control qubit, while $i_{1}$ and $j_{1}$ are input and output for the target qubit. 

The rank and complexity of tensors can be reduced, as is shown in Fig.~\ref{fig1} (d), by exploiting diagonality and merging indices that are always equal into hyper-edges~\cite{b19}. In other words, if the indices are equal, the tensor value is non-zero. For example, $CZ$ is non-zero only when $i_{0}$ equals $j_{0}$ and $i_{1}$ equals $j_{1}$. By exploiting hyper-edges, the rank of $CZ$ is reduced from four to two, decreasing computation during QCS.
Hyper-edges are illustrated with dotted lines.

\begin{figure}[htbp]
\centering
\includegraphics[width=\linewidth]{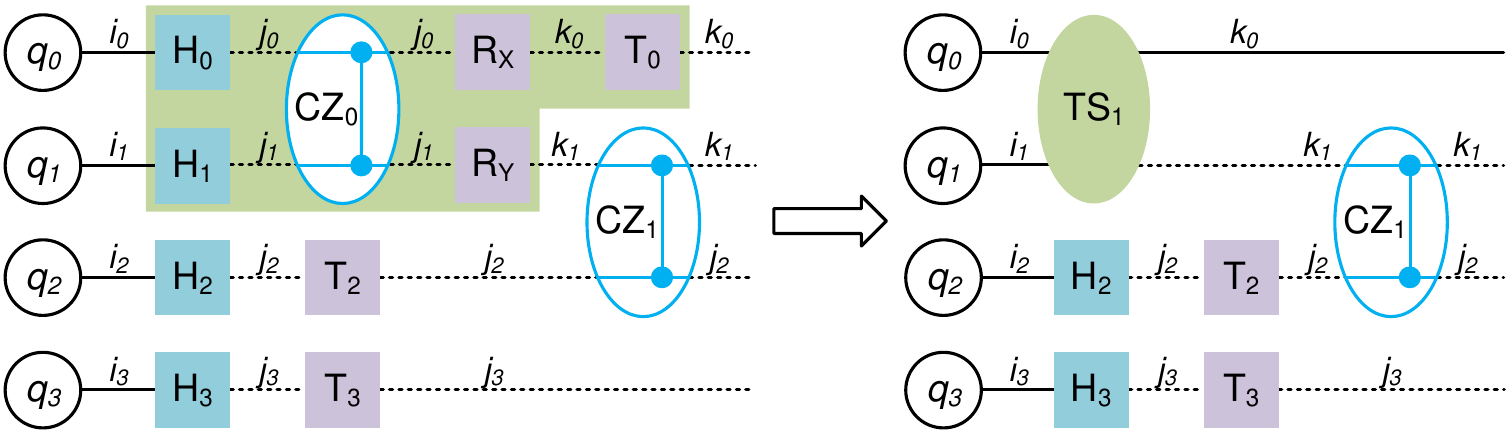}
\caption{The example quantum circuit represented as a tensor network before and after contracting the tensors highlighted in green. $R_{X}$, $R_{Y}$ and $T$ are common single-qubit rotation gates.}
\label{fig2}
\vspace{-5pt}
\end{figure}

A tensor network is a graph of tensors connected through indices. Contraction is the fundamental operations for TN-based QCS; it applies quantum gates to quantum states and calculates the inner product of tensors over shared indices. Quantum circuits inherently exhibit a TN structure. When simulating a quantum circuit as a TN, all the shared indices between the quantum state and gate tensors are contracted, resulting in a single tensor indicated by open indices, such as index $k_{0}$ in Fig.~\ref{fig2}. 
A tensor network before and after contracting the tensors highlighted in green is shown in Fig.~\ref{fig2}, for which hyper-edges have been exploited. Hyper-edges that extend to the open indices (e.g., index $k_{0}$ of $T_{0}$) are not considered as shared indices. The contraction for the highlighted tensors of Fig.~\ref{fig2}, tensor $TS_1$, is mathematically described in Eq.~\ref{eq1}. 

\begin{equation}
TS_{1, i_{0}, i_{1}}^{k_{0}, k_{1}} = \sum_{j_{0}, j_{1}}{H_{0, i_{0}}^{j_{0}} H_{1, i_{1}}^{j_{1}} CZ_{0, j_{0}, j_{1}} R_{X, j_{0}}^{k_{0}} R_{Y, j_{1}}^{k_{1}} T_{0, k_{0}}}
\label{eq1}
\end{equation}

In practice, tensors are rarely contracted all at once. A usual approach involves contracting two tensors at a time. When these tensors are contracted in the sequential order, i.e., qubit by qubit and layer by layer, TN-based QCS essentially mirrors the $\text{Schr\"odinger}$-style simulation. However, as will be detailed in \ref{sec3B}, one of the key advantages of TN lies in its capability to perform tensor contractions in arbitrary orders, which can lead to significant reduction in computation.
For a more in-depth understanding of TN, we recommend consulting~\cite{b19, b20, b21}.

\subsection{Tensors as Tensor Decision Diagrams}


TDDs~\cite{b26} are DAGs that serves as a compact representation for tensors. They can be considered as a variant of reduced-ordered binary decision diagrams (ROBDD)~\cite{b27}, with directed edges weighted in complex numbers, essentially functioning as a mapping from binary index values to corresponding complex tensor values.
To evaluate the tensor value for certain index values, we start from the root edge and traverse the TDD. At each node, we take the left (right) edge to traverse to a child node if the corresponding index is 0 (1). This process continues until we reach a leaf node, and the tensor value is the product of all the edge weights that we have traversed.

\begin{figure}[ht]
\centering
\includegraphics[width=\linewidth]{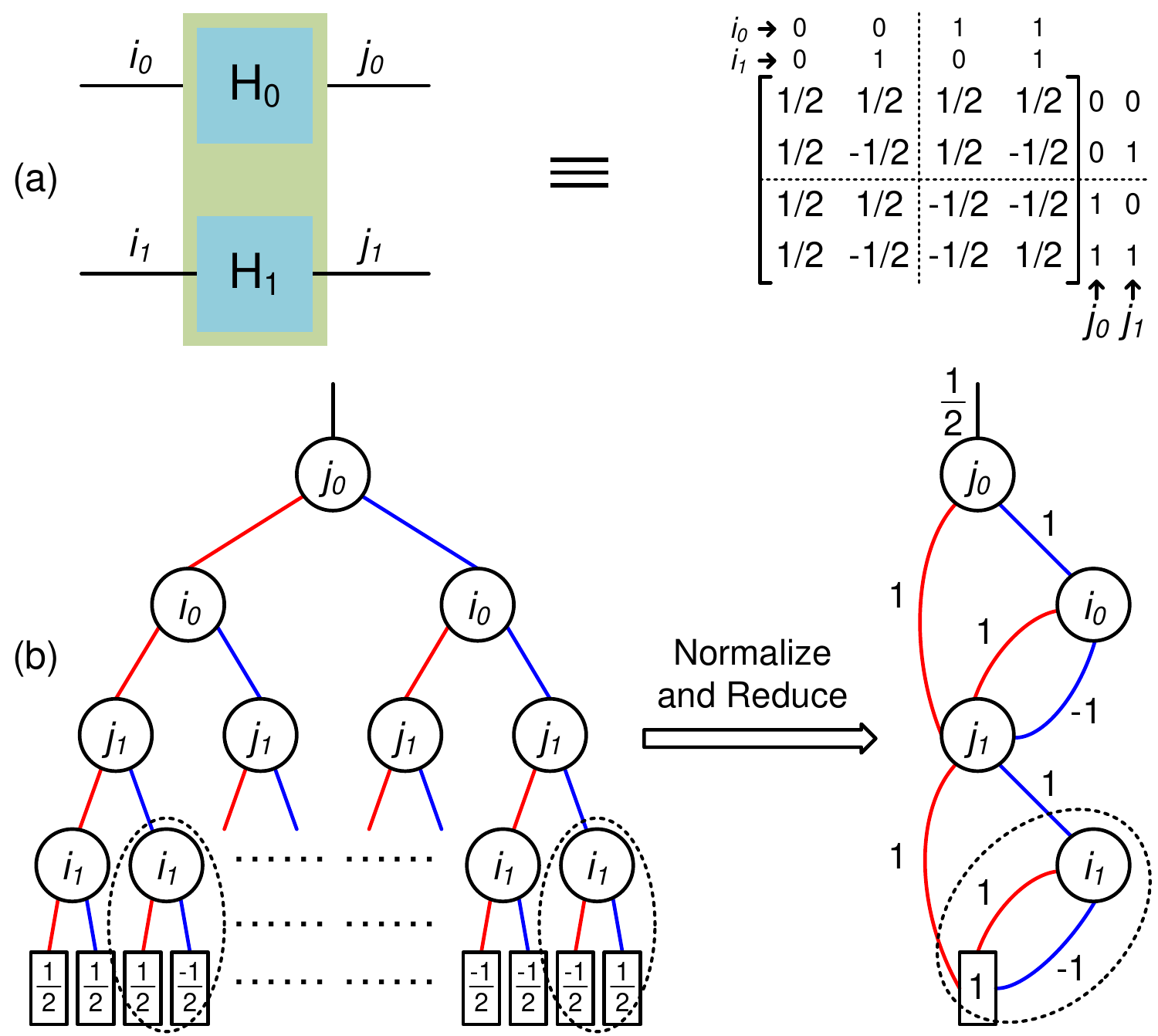}
\caption{(a) The tensor $H \otimes H$ and its array; (b) Its TDD before and after normalization and reduction. A red (blue) edge indicates an index of 0 (1).}
\label{fig3}
\end{figure}

Fig.~\ref{fig3} shows how the tensor product of two $H$ gates is converted to TDD. The array (Fig.~\ref{fig3} (a) right) is first Shannon-expanded into a binary tree (Fig.~\ref{fig3} (b) left). Then, the nodes are normalized and reduced level by level, starting from the leaves and moving up to the root. At each node, the normalization is performed by dividing the two output edge weights by the one with the larger magnitude. The resulting divider is then propagated up to the node's incoming edge and multiplied with the corresponding weight. After normalization, reduction begins by checking if the normalized node already exists in the unique table, a hash table for storing nodes uniquely. If it does, the node is linked to its parent node. Otherwise, a new node is created in the unique table. For example, after normalization and reduction, circled nodes in Fig.~\ref{fig3} (b) (left) are merged into the circled node in Fig.~\ref{fig3} (b) (right). This reduces the number of nodes for  $H \otimes H$ from 16 to 4. TDD supports all the essential tensor operations utilized in QCS through recursion, including contraction and addition. For a comprehensive understanding of TDD, we recommend consulting~\cite{b26}.

\section{Tensor Network Optimizations}\label{sec3}

\subsection{Efficient Rank Simplification with Tetris}

In the context of TNs, rank simplification is an effective preprocessing step~\cite{b21} that reduces the number of gates to be simulated by consolidating tensors in a local manner.
Specifically, it involves identifying groups of adjacent tensors in a TN and contracting the tensors in each group, under the constraint that for each group, the rank of the resulting tensor does not exceed the ranks of any contracted tensors. Unfavorably, a TN is typically supplied as a list of tensors without locality information, and rank simplification therefore entails a brute-force iteration through the tensor list. For each tensor, the rest tensors are further iterated and checked against it one by one, and are contracted with it only if they are adjacent and follow the constraint.
This approach results in a computational complexity of $O(T^2)$, where $T$ is the number of tensors.
However, DD-based quantum states can exhibit linear complexity $O(n)$, where $n$ is the number of qubits. With at most $T$ gates to be applied, simulating the rank-simplified circuit can results in a complexity of $O(nT)$.
Considering that $T$ is usually much larger than $n$, the rank simplification preprocessing could take longer than the simulation itself.

We notice that the brute-force approach mentioned above assumes that TNs have arbitrary structures, which is not always the case and can result in unnecessary overhead. We contend that the sequential layer-by-layer structure of quantum circuits can be leveraged to minimize rank simplification overheads. In sequential structures, rank simplification is restricted to gates that are from adjacent layers and share qubits.
Using this insight, we developed Tetris, an algorithm designed to achieve rank simplification in quantum circuits with only linear complexity, $O(T)$.
The name of the algorithm is inspired by the popular game Tetris. In Tetris, various-shaped bricks, known as tetrominoes, descend and stack on the ground. When rows of bricks are completely filled, they merge and disappear. Intriguingly, our algorithm follows a similar pattern.

\begin{figure}[ht]
\centering
\includegraphics[width=\linewidth]{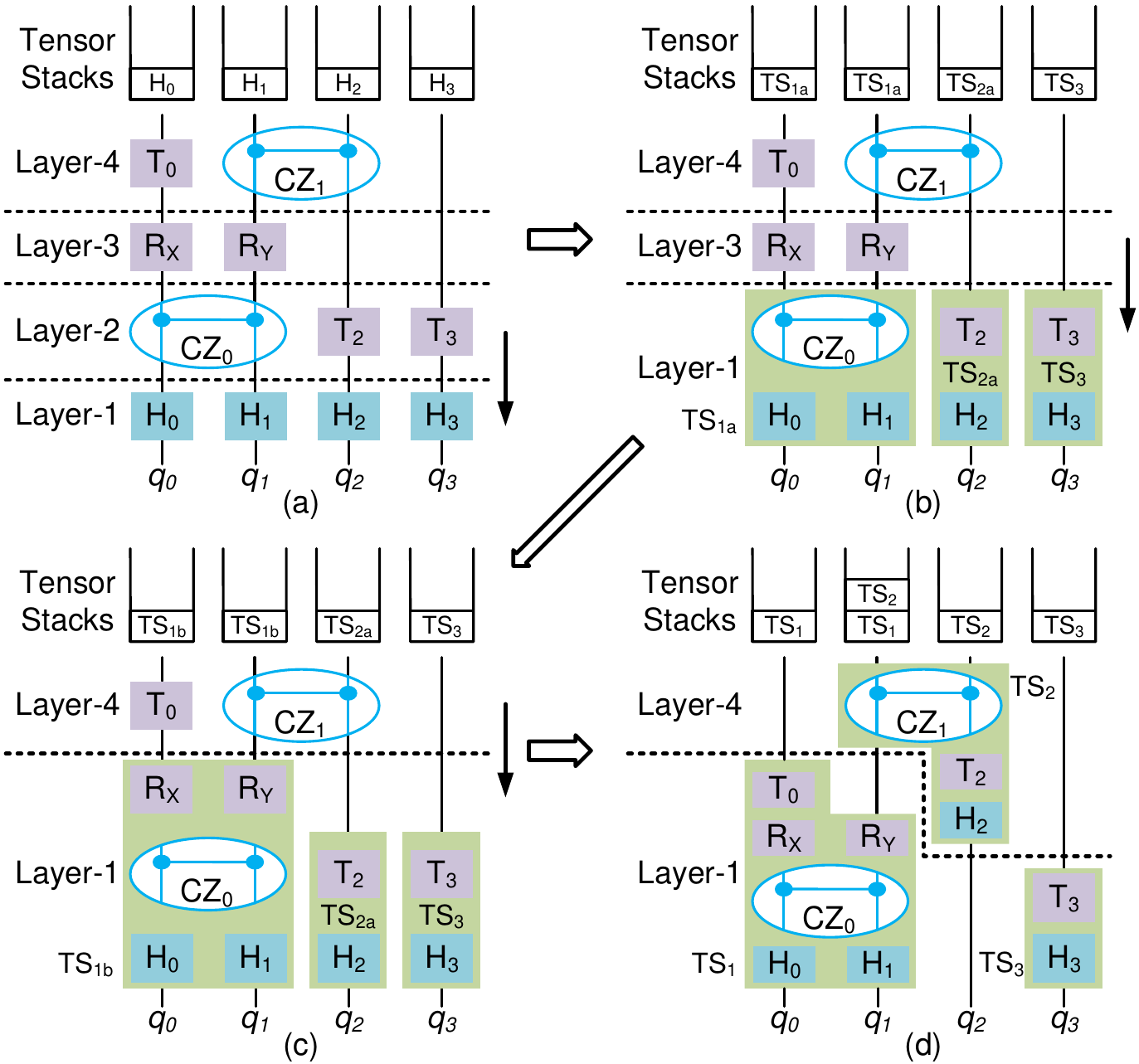}
\caption{Applying our Tetris algorithm to the example circuit of Fig.~\ref{fig2}. Panels (a) to (d) depict the transition from the original circuit to the simplified version as we proceed through a layer-by-layer process, commencing from layer 2.}
\label{fig4}
\end{figure}

Fig.~\ref{fig4} illustrates the proposed gradual rank simplification algorithm for the previously introduced example circuit, with quantum gates resembling tetrominoes.
It initiates by establishing tensor stacks shown in Fig.~\ref{fig4} (a), one for each qubit line, containing consolidated tensors along their respective qubit lines. Subsequently, the circuit is systematically examined, progressing through its layers in a sequential manner.
Within each layer, the gates are handled as if they were descending along the qubit lines, akin to tetrominoes in the game of Tetris. Every gate is examined in relation to the top tensors in the stacks
corresponding to its associated qubits. If there is qubit sharing between the gate and the top tensors, and the rank simplification constraint is met, the gate is consolidated with the top tensors via a contraction, and the outcome tensor replaces the top tensors. Otherwise, the gate is pushed to the top positions in the stacks. For example, $CZ_{1}$ and $TS_{2a}$ in Fig.~\ref{fig4} (c) can be consolidated into $TS_{2}$ in Fig.~\ref{fig4} (d), while  $TS_{1}$ and $TS_{2}$ are not further consolidated, which violates the constraint. After the last layer is considered, the rank-simplified circuit is retrieved by popping the bottom tensors from the stacks layer-by-layer, with duplicated ones discarded.
For the example circuit, Tetris effectively reduces the number of gate tensors from 11 to 3.

\subsection{Execution of Near-Optimal Contraction Order}\label{sec3B}

\begin{figure}[ht]
\centering
\includegraphics[width=\linewidth]{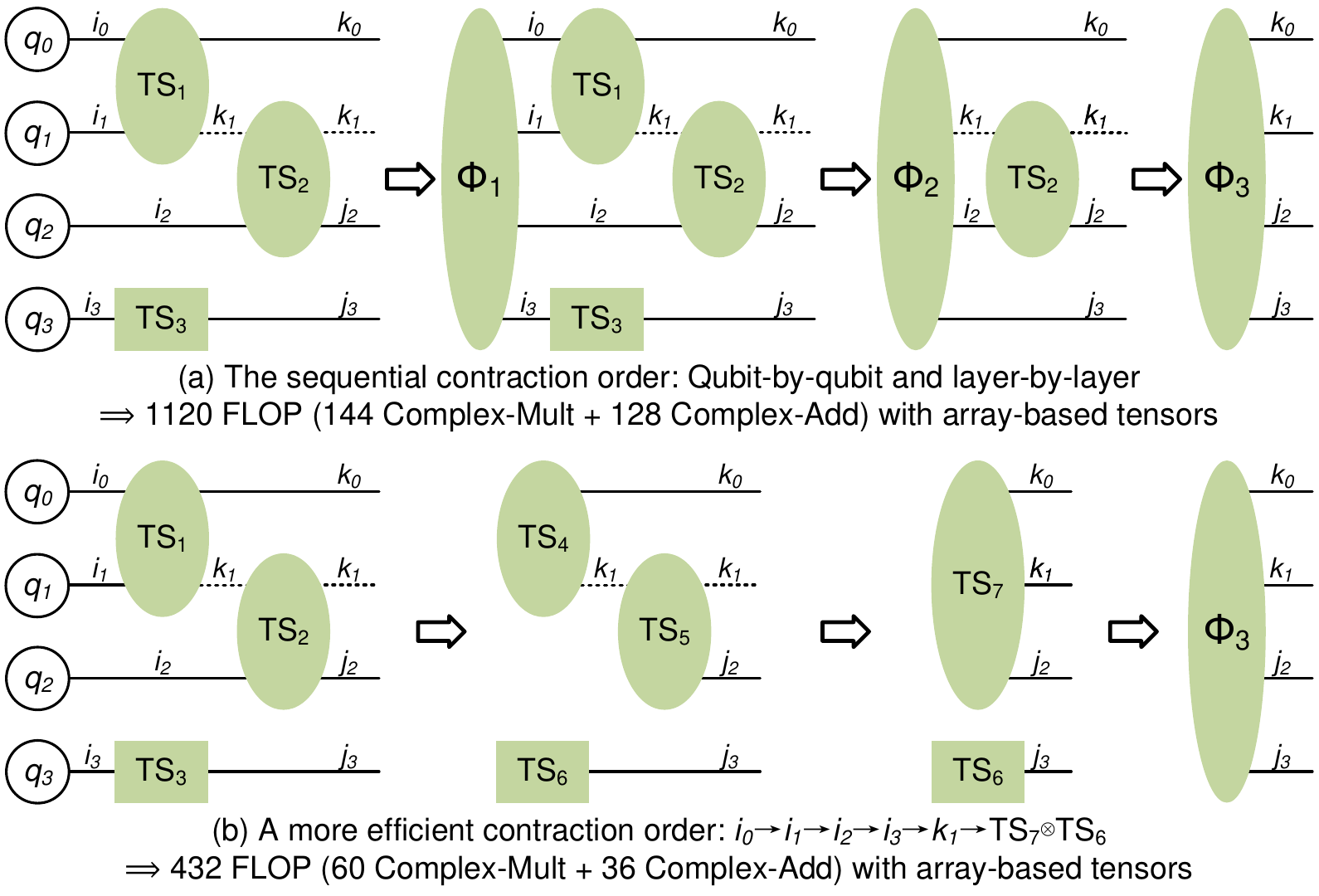}
\caption{Different contraction orders for the rank-simplified example circuit.}
\label{fig5}
\end{figure}

In a TN, the order in which tensors are contracted can vary, but not all ordering choices yield the same level of efficiency.
Fig.~\ref{fig5} shows two different contraction orders for the rank-simplified example circuit. In Fig.~\ref{fig5} (a), the sequential order is applied, and immediately after the first layer, intermediate tensor rank grows to four. On the other hand, Fig.~\ref{fig5} (b) shows a more efficient order where contractions that results in larger tensors are delayed as much as possible, and the rank of intermediate tensor only grows to four at the last step. With the more efficient order, the number of floating point operations (FLOP) using arrays to simulate the rank-simplified example circuit is reduced from 1,120 to only 432.

\begin{figure}[ht]
\centering
\includegraphics[width=\linewidth]{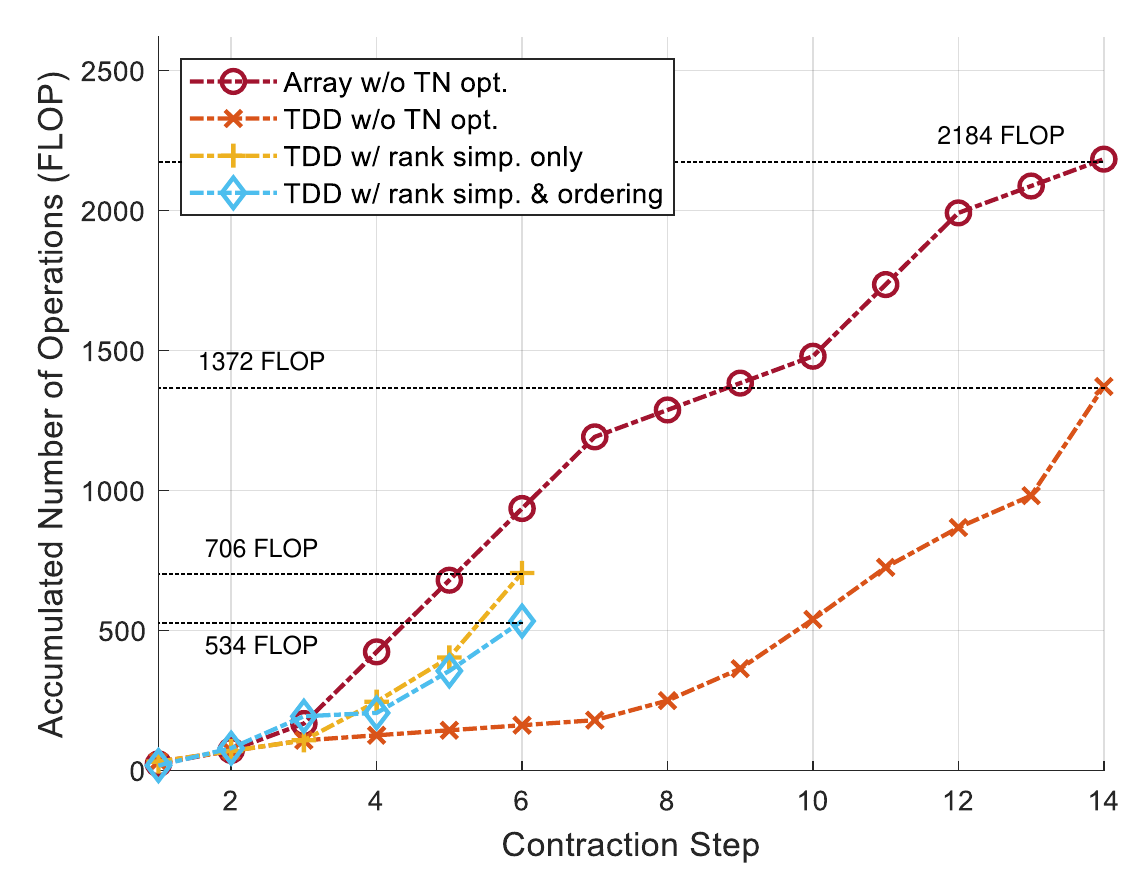}
\caption{FLOP progressions for the example circuit with different methods.}
\label{fig6}
\end{figure}

As the rank of a tensor aligns with the number of levels in its TDD representation, we hypothesize that optimizing the order of contractions can also lead to complexity reduction in TDD-based TNs. This reduction occurs by minimizing the number of levels and, consequently, the number of nodes within intermediate TDDs. Taking advantage of this opportunity, we introduce the concept of executing near-optimal contraction orders for TDD-based TNs. This represents the first application of such optimization techniques to decision diagrams that are compatible with TN formalism. In our practical implementation, we rely on Cotengra~\cite{b21} as the backend tool to search for these near-optimal TN contraction orders.
Additionally, we combine contraction ordering with TDD and our Tetris algorithm to synergistically reduce the complexity of quantum circuit simulation even further.
Fig.~\ref{fig6} illustrates how TDDs decrease the number of FLOPs in array-based QCS from 2,184 to 1,372. Application of the Tetris algorithm goes a step further, reducing the number of contraction steps from 14 to 6 and decreasing FLOPs from 1,372 to 706. Ultimately, when employing the contraction order in Fig.~\ref{fig5} (b), the FLOP count for TDD-based contraction is minimized to just 534. In summary, the proposed approach results in a 76\% reduction in computation for the example circuit.

\section{Tensor Decision Diagram Optimizations}\label{sec4}

To simulate the final quantum state, the TN resulting from the above optimizations must be contracted globally. To exploit redundancy and reduce computation, we encode tensors as TDDs for the global contraction. In this section, we present our sequence of TDD optimizations, implemented in C++. The prior TDD work, PyTDD~\cite{b26}, is based on Python. Our results show that the transition from Python to C++ alone yields a speedup of $\sim$$2.5\times$ on Google RQCs.

\subsection{Edge-centric Data Structures for TDD}

\begin{figure}[ht]
\centering
\includegraphics[width=\linewidth]{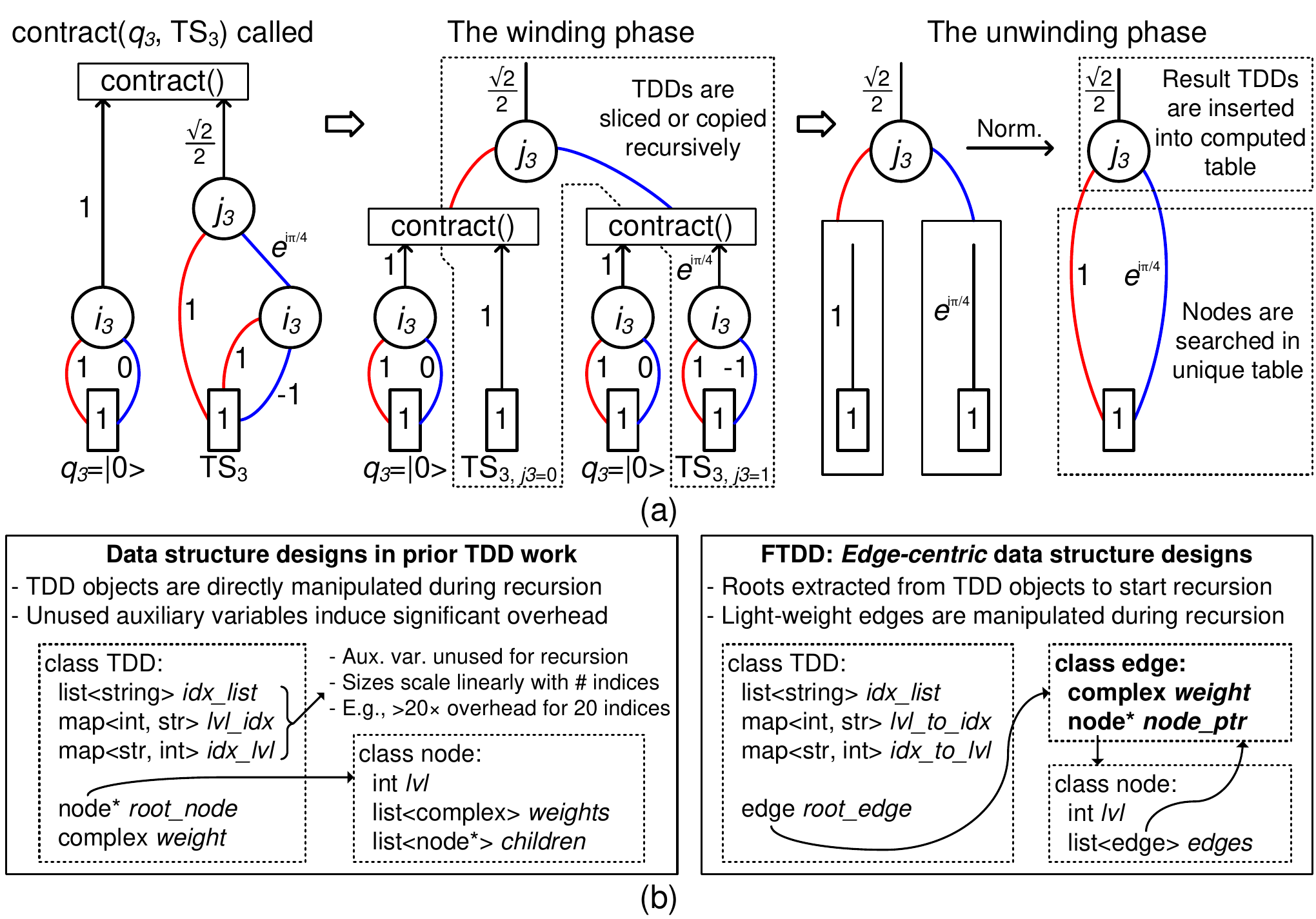}
\caption{(a) Illustration of the recursive process in TDD contraction using $q_{3}$ and $TS_{3}$ from the rank-simplified example circuit; (b) Comparison of data structure designs between the prior TDD work and FTDD.}
\label{fig7}
\end{figure}

TDD contraction begins with a one-time preparation of control variables, followed by a recursive process outlined in Fig.~\ref{fig7} (a).
During recursion, frequent manipulation of intermediate TDDs is the primary source of latency. Therefore, optimizing the underlying data structure designs for TDDs is crucial.
Fig.~\ref{fig7} (b) (left) shows the encoding of TDDs in TDD objects, as in PyTDD~\cite{b26}. Each TDD object contains a root node, incoming edge weight, and auxiliary variables for managing tensor indices ($idx\_list$) and index-to-level correspondences ($idx\_lvl$ and $lvl\_idx$). We notice that while auxiliary variables are essential for preparing the control variables, they become unnecessary during recursion and possess cumbersome sizes that scale linearly with the number of indices.

With that observation, we identify that a major inefficiency is the direct use of TDD objects during recursion.
As Fig.~\ref{fig7} (a) illustrates, the recursion involves two phases: winding and unwinding. During winding, intermediate TDDs are sliced into child TDDs to incur downstream calls, where TDDs without nodes at a level are copied. For example, the $TS_{3}$ TDD is split into two children at level $j_{3}$, while $q_{3}$ TDD is copied since it has no nodes at that level.
However, the massive auxiliary variables are also unnecessarily copied, inducing significant overheads. 
During unwinding, recursive calls return level by level from bottom up. For each return, intermediate result TDDs are inserted into the computed table, which stores computed TDDs to be reused by later calls with the same input operands. However, for this purpose, only the root node and incoming edge weights are required.
Using TDD objects, useless auxiliary variables are also stored in the computed table, inducing undesired overheads for the access of the computed table and memory size.

To address the overhead issues,
we propose edge-centric data structures for TDD. Shown in Fig.~\ref{fig7} (b) (right), we introduce the \textit{edge} objects in FTDD to replace TDD objects throughout recursion. To initiate recursion, root edges are extracted from operand TDD objects to encode TDDs. During recursion, only lightweight edge objects are manipulated, and computed tables only store edge objects, completely removing the involvement of auxiliary variables during recursion. Additionally, a node object is simplified to store only one list for edges, instead of two lists for output edge weights and children nodes, reducing unique table access overhead.

\subsection{Employment of Key BDD Optimizations}

While specialized for tensors, TDD still shares common building blocks with the other decision diagrams. Here, we explore the connection between TDD and the wealth of prior DD research for the first time. We find that TDD can effectively benefit from key BDD ~\cite{b30} optimizations that are essential for building efficient unique tables and computed tables. However, it is worth noting that certain BDD optimizations such as \textit{complement edges} and \textit{standard triple} are specific to Boolean functions, and do not apply to TDD.

FTDD incorporates three BDD optimizations. First, FTDD adopts \textit{merged-DAG unique table}, where the memory for storing all TDD nodes is augmented with hashing support and can be repurposed as the unique table, eliminating memory overheads from storing duplicated nodes in a separate unique table.
Second, FTDD supports \textit{garbage collection} to manage memory efficiently. It counts references to each node and frees up unused memory. After each contraction, the reference counts for all nodes are updated, and when the unique table's node count exceeds a certain limit, unreferenced nodes are identified and removed.
Lastly, FTDD employs \textit{caching computed table}, which simplifies computed tables from size-varying hash tables into fixed-size caches addressed by hashes. Upon cache miss, incoming data replaces existing data instead of being appended. This approach minimizes computed table overheads by controlling its size to match actual needs.

\begin{algorithm}
\caption{FTDD tensor network contraction}\label{alg1}
\begin{algorithmic}

\Require $unique\_table$, $tensor\_list$, $order$
\Ensure $tdd\_res$
\vspace{3pt}

\State{$tdd\_list \gets$  empty list of TDD}
\vspace{3pt}
\For{$tensor$ in $tensor\_list$} \Comment{Convert tensors to TDDs}
    \State $tdd\_tmp \gets$ tensor\_to\_tdd($tensor$)
    \State $tdd\_list$.append($tdd\_tmp$)
    \State $unique\_table$.inc\_ref\_cnt($tdd\_tmp$.root\_edge)
\EndFor
\vspace{3pt}
\For{$pair$ in $order$} \Comment{Contraction with the given order}
    \State $tdd\_a \gets$ $tdd\_list$[$pair$.first]
    \State $tdd\_b \gets$ $tdd\_list$[$pair$.second]
    \State $tdd\_c \gets$ contract($tdd\_a$, $tdd\_b$)
    \vspace{3pt}
    \State $unique\_table$.dec\_ref\_cnt($tdd\_a$.root\_edge)
    \State $unique\_table$.dec\_ref\_cnt($tdd\_b$.root\_edge)
    \State $unique\_table$.inc\_ref\_cnt($tdd\_c$.root\_edge)
    \vspace{3pt}
    \State $tdd\_list$.remove($tdd\_a$)
    \State $tdd\_list$.remove($tdd\_b$)
    \State $tdd\_list$.append($tdd\_c$)
    \vspace{3pt}
    \If {$unique\_table$.size $>$ $unique\_table$.size\_limit  }
        \State $unique\_table$.garbage\_collection() 
    \EndIf
\EndFor
\vspace{3pt}
\State $tdd\_res \gets$ $tdd\_list$[0]

\end{algorithmic}
\end{algorithm}

Combining aforementioned TN and TDD optimizations, Algorithm~\ref{alg1} illustrates the contraction of a tensor network in FTDD. The process begins with the initialization of a TDD list by converting tensors in a TN into TDDs. As the TDDs are newly created, reference counts of nodes are incremented for each of them. Then, the TDDs are contracted based on a specified order presented as a sequence of pairs. Each pair denotes the locations of tensors to be contracted in the dynamically updated TDD list. Following each contraction, the operand TDDs are removed from the list, and the result TDD is appended to it. Since the removed TDDs are essentially de-referenced, the reference counts of their nodes can be safely decremented. For the newly created result TDD, reference counts of its nodes are incremented. Garbage collection is triggered if the unique table size exceeds a user-defined limit. Throughout the process, TDD list length steadily decreases until reaching one after the final contraction, where the sole remaining TDD represents the QCS result.

\section{Evaluation}

\subsection{Setup}

All our experiments were run on a single thread on a workstation with Intel® Xeon® Gold 6254 CPUs and 240 GiB DRAM running Ubuntu 20.04 LTS. Cotengra is configured to conduct 512 searches for each TN.

\subsection{Benchmarks}

For comprehensive benchmarking, we consider a wide range of quantum circuits from MQT Bench~\cite{b34} and Google~\cite{b29}, including both circuits that exhibit rich redundancy, such as GHZ, graph state, and quantum Fourier transform (QFT), and hard-to-simulate circuits like entangled-QFT and Google RQCs.
For QFT, we follow~\cite{b26} and conduct its unitary simulation, which treats the input qubits as open indices and results in rank-2$n$ tensors for $n$ qubits. Additionally, we benchmark quantum approximate optimization algorithm (QAOA) and variational quantum eigensolver (VQE), as they are representative of noisy intermediate-scale quantum (NISQ)~\cite{b7} applications.

\subsection{Unique Table and Computed Table Statistics}

The incorporation of garbage collection and caching computed tables in FTDD represents significant modifications to the original PyTDD design. To ensure that these changes do not adversely impact the hit rates of unique table and computed tables, potentially leading to performance degradation, we verified the table statistics of FTDD against PyTDD on Google RQCs. Table.~\ref{tab3} presents the hit rates of unique table, contraction computed table, and addition computed table for PyTDD and FTDD across four 16-qubit Google RQCs, executing TN contractions in sequential orders. The results indicate a negligible difference in hit rates between PyTDD and FTDD. Intriguingly, the unique table hit rates of FTDD are occasionally slightly higher than those of PyTDD.

\begin{table*}[htbp]
\caption{Unique Table and Computed Table Hit Rates of PyTDD and FTDD on 16-qubit Google RQCs}
\begin{center}
\begin{tabular}{c|c|c|c|c|c|c}

\hline
Circuit                     & \multicolumn{2}{c|}{Unique Table Hit Rate}    & \multicolumn{2}{c|}{Contraction Computed Table Hit Rate}  & \multicolumn{2}{c}{Addition Computed Table Hit Rate} \\
\cline{2-7}                                                                                                              
                            & PyTDD     & FTDD                              & PyTDD     & FTDD                                          & PyTDD     & FTDD \\
\hline
\textit{inst\_4x4\_10\_8}   & 27.52\%   & 27.82\%                           & 29.40\%   & 29.38\%                                       & 5.52\%    & 5.46\% \\
\textit{inst\_4x4\_12\_8}   & 8.06\%    & 8.14\%                            & 35.69\%   & 35.68\%                                       & 1.77\%    & 1.72\% \\
\textit{inst\_4x4\_14\_8}   & 5.57\%    & 5.62\%                            & 30.41\%   & 30.41\%                                       & 1.17\%    & 1.14\% \\
\textit{inst\_4x4\_16\_8}   & 5.52\%    & 5.55\%                            & 28.58\%   & 28.58\%                                       & 0.80\%    & 0.78\% \\
\hline

\end{tabular}
\label{tab3}
\end{center}
\end{table*}

\subsection{Ranks Simplification Overhead with Tetris}

Table~\ref{tab1} shows Tetris runtime for all the benchmarks. With $O(T)$ complexity, Tetris only induces $4\%$ runtime overhead on average when compared to FTDD runtime with TN optimizations, demonstrating its efficiency. Notably, for 20-qubit Google RQCs, Tetris has a negligible $0.05\%$ runtime overhead.

\begin{table*}[htbp]
\caption{Comparison between FTDD and State-of-the-art Array-based and DD-based QCS Methods
\vspace{-5pt}
}

\begin{center}
\resizebox{\linewidth}{!}{
\begin{tabular}{c|c|c|c|c|c|c|c|c|c|c}

\hline
\multicolumn{3}{c|}{Benchmarks}                         & \multicolumn{2}{c|}{Google TensorNetwork} & QMDD          & PyTDD     & \multicolumn{4}{c}{FTDD} \\
\cline{1-5}\cline{8-11}
Circuit & \# qubits  & \# gates   & seq. time (s) & opt. time (s)             & time (s)      & time (s)  & seq. time (s) & Tetris time (s)  & opt. time (s) & final \# nodes \\
\hline
GHZ                             & 25        & 25        & 16.2          & 0.31        & 0.044         & 0.032     & 0.037         & 0.002     & 0.03    & 50 \\
GHZ                             & 26        & 26        & 32.86         & 0.66        & 0.044         & 0.034     & 0.037         & 0.002     & 0.03    & 52 \\
GHZ                             & 27        & 27        & 67.54         & 1.43         & 0.044         & 0.036     & 0.039         & 0.002     & 0.03    & 54 \\
GHZ                             & 28        & 28        & 140.95        & 2.29        & 0.044         & 0.038     & 0.039         & 0.002     & 0.031   & 56 \\
\hline\hline
Graph state                     & 25        & 50        & 28.02         & 0.31        & 0.075         & 0.42      & 0.081         & 0.001     & 0.038   & 807 \\
Graph state                     & 26        & 52        & 57.53         & 0.61        & 0.076         & 0.47      & 0.089         & 0.001     & 0.037   & 835 \\
Graph state                     & 27        & 54        & 117.6         & 1.39         & 0.077         & 0.3       & 0.073         & 0.002     & 0.036   & 663 \\
Graph state                     & 28        & 56        & 251.47        & 2.58        & 0.079         & 0.48      & 0.095         & 0.001     & 0.041   & 1359 \\
\hline\hline
QFT (unitary)                   & 13        & 91        & 56.4          & 6.0          & 1.13          & 6.68      & 0.67          & 0.004     & 0.15    & 16383 \\
QFT (unitary)                   & 14        & 105       & 261.68        & 22.34       & 4.21          & 14.11     & 1.35          & 0.005     & 0.4     & 32767 \\
QFT (unitary)                   & 15        & 120       & out of mem.          & out of mem.                      & 16.51         & 28.79     & 3.26          & 0.007     & 0.86    & 65535 \\
QFT (unitary)                   & 16        & 136       & out of mem.          & out of mem.                      & 67.88         & 55.46     & 9.04          & 0.008     & 2.64     & 131071 \\
\hline\hline
Entangled-QFT                   & 14        & 126       & 0.049         & 0.025       & 1.93          & 4.01      & 0.27          & 0.008     & 0.074    & 16384 \\
Entangled-QFT                   & 15        & 142       & 0.07          & 0.034       & 18.66         & 7.89      & 0.49          & 0.011     & 0.11     & 32768 \\
Entangled-QFT                   & 16        & 160       & 0.11          & 0.053       & 166.98        & 15.37     & 0.92          & 0.014     & 0.19     & 65536 \\
Entangled-QFT                   & 17        & 178       & 0.21          & 0.1          & 1461.96       & 29.33     & 1.79          & 0.017     & 0.37     & 131072 \\
\hline\hline
QAOA                            & 12        & 60        & 0.019         & 0.005       & seg. fault    & 5.35      & 0.3           & 0.002     & 0.1      & 3078 \\
QAOA                            & 13        & 65        & 0.02          & 0.006       & seg. fault    & 12.23     & 0.65          & 0.002     & 0.22    & 6203 \\
QAOA                            & 14        & 70        & 0.024         & 0.005       & seg. fault    & 1.84      & 0.14          & 0.002     & 0.014   & 512 \\
QAOA                            & 15        & 75        & 0.036         & 0.007       & seg. fault    & 47.88     & 2.66          & 0.002     & 0.8      & 24709 \\
\hline\hline
VQE                             & 16        & 78        & 0.05          & 0.007       & 1.61          & div. zero & 0.36          & 0.002     & 0.33    & 36820 \\
VQE                             & 17        & 83        & 0.091         & 0.008       & 0.43          & div. zero & 0.033         & 0.003     & 0.02    & 837 \\
VQE                             & 18        & 88        & 0.18          & 0.008       & 6.85          & div. zero & 0.062         & 0.003     & 0.049    & 5085 \\
VQE                             & 19        & 93        & 0.34          & 0.011       & 0.49          & div. zero & 0.054         & 0.003     & 0.036    & 853 \\
\hline\hline
RQC (\textit{inst\_4x5\_10\_8}) & 20        & 145       & 1.32          & 0.014       & 49.08         & 828.12    & 43.7          & 0.004     & 3.23    & 454658 \\
RQC (\textit{inst\_4x5\_12\_8}) & 20        & 168       & 1.46          & 0.023       & 416.95        & 2091.5    & 106.76        & 0.005     & 13.58   & 1048576 \\
RQC (\textit{inst\_4x5\_14\_8}) & 20        & 192       & 1.78          & 0.055       & 790.89        & 2863.04   & 153.08        & 0.006     & 37.41   & 1048576 \\
RQC (\textit{inst\_4x5\_16\_8}) & 20        & 215       & 2.13          & 0.081        & 1274.46       & $>$3600     & 183.54        & 0.008     & 41.15    & 1048576 \\
\hline

\multicolumn{11}{c}{seq. time - Simulating original circuits in sequential contraction orders, opt. time - Best runtime when considering all possible combinations of TN optimizations} \\

\end{tabular}
}
\label{tab1}
\end{center}
\vspace{-10pt}
\end{table*}

\subsection{Comparison Against DD-based QCS Approaches}

\begin{figure}[htbp]
\centering
\includegraphics[width=0.9\linewidth]{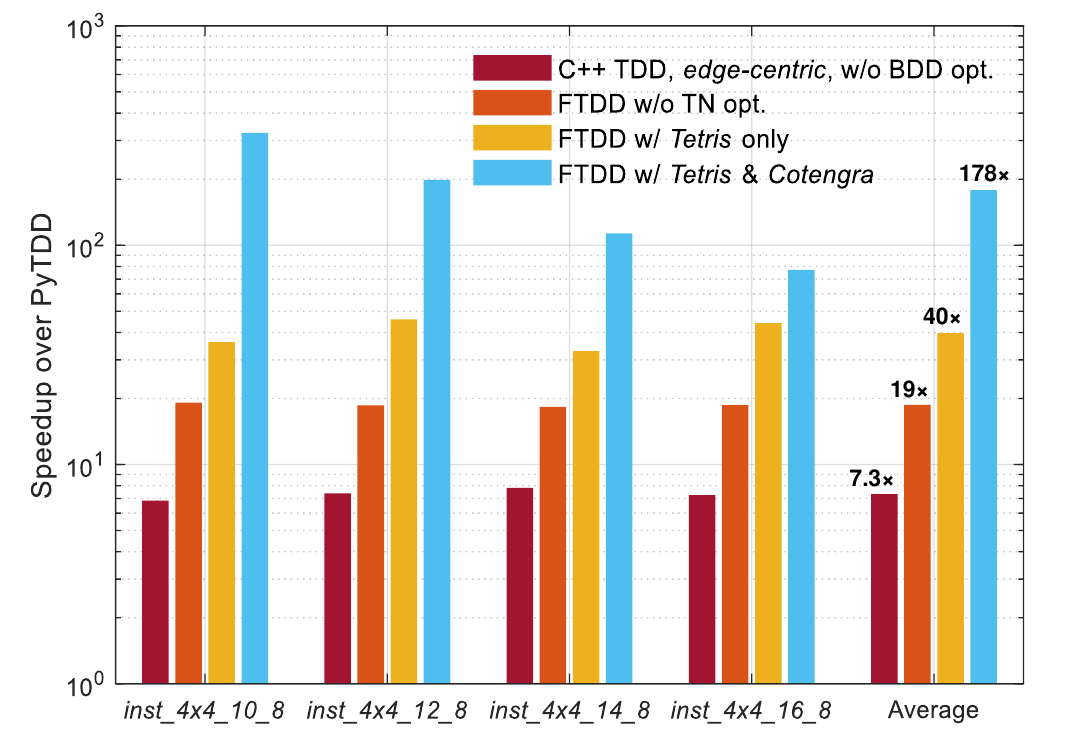}
\caption{FTDD speedup with successive optimizations on 16-qubit RQCs.}
\label{fig9}
\end{figure}

For DD-based approaches, we first experiment with the speedup of FTDD over PyTDD~\cite{b26} using incremental combinations of FTDD optimizations, on four 16-qubit Google RQCs of different depths. Fig.~\ref{fig9} shows the results of this experiment.
As Section \ref{sec4} mentions, simply transpiling Python to C++ only gives $\sim$$2.5\times$ speedup on average. With edge-centric data structure designs, the average speedup for the 16-qubit RQCs is increased to $7.3\times$.
Employing key BDD optimizations further boosts the average performance to be $19\times$ faster.
With Tetris, 75\% of the gates can be reduced on average, marching the speedup to $40\times$.
Finally, executing near-optimal contraction orders found through Cotengra for the rank-simplified circuits propells FTDD performance to an astounding average speedup of $178\times$ compared to PyTDD.
We further compare FTDD against PyTDD for all benchmarks. Table~\ref{tab1} shows the runtime of FTDD and PyTDD across the benchmarks. Table~\ref{tab2} shows the average speedup of FTDD over PyTDD for each type of circuits.
With TN optimizations, FTDD achieves speedup over PyTDD on all the benchmarks. Remarkably, on the hardest benchmark, 20-qubit RQCs, FTDD exhibits $143.7\times$ speedup over PyTDD with the TN optimizations applied.
Optimized FTDD also shows an impressive $74.8\times$ speedup over PyTDD on QAOA, a NISQ application. In addition, PyTDD ran into division-by-zero errors when simulating VQE, indicating that FTDD may offer improved robustness.

\begin{table}[htbp]
\caption{
Average Speedup of FTDD over PyTDD and QMDD
\vspace{-5pt}
}

\begin{center}
\resizebox{\linewidth}{!}{
\begin{tabular}{c|c|c|c|c}

\hline
Benchmarks              & \multicolumn{2}{c|}{FTDD seq. time}  & \multicolumn{2}{c}{FTDD opt. time} \\
\cline{2-5}                                                                                                              
                        & vs. PyTDD & vs. QMDD                  & vs. PyTDD     & vs. QMDD \\
\hline
GHZ                     & $0.9\times$   & $1.2\times$           & $1.2\times$   & $1.5\times$ \\
Graph state             & $4.9\times$   & $0.9\times$           & $11.0\times$  & $2.0\times$ \\
QFT (unitary)           & $8.9\times$   & $4.4\times$           & $33.2\times$  & $15.7\times$ \\
Ent. QFT                & $16.2\times$  & $261.7\times$         & $72.0\times$  & $1248.5\times$ \\
QAOA                    & $17.1\times$  & N/A                   & $74.8\times$  & N/A \\
VQE                     & N/A           & $34.0\times$          & N/A           & $44.6\times$ \\
RQC                     & $19.2\times$  & $4.3\times$           & $143.7\times$ & $24.5\times$ \\
\hline

\end{tabular}
}
\label{tab2}
\end{center}
\vspace{-10pt}
\end{table}

Then, we compare FTDD with QMDD~\cite{b25, b37}, the state-of-the-art DD for QCS. 
We configure QMDD to use its \textit{qasm\_simulator}, ensuring that its runtime does not involve the final DD-to-array conversion. 
Additionally, we allow QMDD to gain speedup using its quantum state approximation techniques~\cite{b38}, but with fidelity no lower than 99.9999\% for fairness, since that is the lowest fidelity FTDD achieves when being verified against IBM Qiskit simulation on the benchmarks.
Table ~\ref{tab1} presents QMDD runtime across all benchmarks, while Table~\ref{tab2} shows the average speedup of FTDD over QMDD for each type of circuits. FTDD achieves speedup over QMDD across all the benchmarks when TN optimizations are applied, establishing FTDD the fastest DD framework for QCS to date.
For 20-qubit Google RQCs, FTDD exhibits $4.3\times$ speedup over QMDD even without leveraging TN optimizations. With TN optimizations applied, FTDD achieves $24.5\times$ speedup over QMDD.
Furthermore, FTDD shows an impressive $44.6\times$ speedup over QMDD on VQE. Notably, QMDD ran into segmentation faults when simulating QAOA, suggesting that FTDD may also offer better robustness.

For circuits such as GHZ and graph state, FTDD demonstrates only a moderate speedup over PyTDD and QMDD. This is primarily due to the substantial redundancy presented in these circuits. The final number of nodes in Table~\ref{tab1} indicates that their quantum states tend to display near-linear complexity. Consequently, the runtime can be influenced more by the overhead from auxiliary steps, such as the initial conversion of gate tensors from arrays to TDDs, rather than the execution of TDD contractions. Importantly, these auxiliary steps do not scale with the proposed optimizations.

\subsection{Comparison Against Array-based TN Approach}

Finally, we compare FTDD with GTN~\cite{b35}, the state-of-the-art array-based TN simulator. 
We configured GTN to use PyTorch as the backend and limited it to also use only a single thread for a fair comparison. However, it is important to note that we are not able to disable PyTorch's vector execution, as it is compiled into the software.
As Tetris and Cotengra are also applicable to array-based TNs, we also present GTN results both with and without the TN optimizations.

In the cases of hard-to-simulate circuits like entangled-QFT and RQCs, Table~\ref{tab1} shows that FTDD can have a longer runtime than GTN. One reason is that GTN takes advantage of vector execution, while FTDD does not employ any concurrency. Moreover, the computational cost of manipulating a TDD node and its edges is higher than simply computing an array entry. For hard-to-simulate circuits, TDDs cease to exploit redundancy as circuits reach certain depths and eventually expand to full binary trees with exponentially many nodes, the same complexity as arrays, leading to overall performance degradation.
However, FTDD demonstrates superiority for the redundancy-rich circuits. As Table~\ref{tab1} shows, the final FTDD node counts for final quantum states of GHZ and graph state circuits demonstrate that they only exhibit near-linear complexity when using FTDD. For the final quantum states of QFT unitary simulation, array-based simulation imposes $O(2^{2n})$ complexity. Using FTDD, this complexity
can be exponentially reduced to only $O(2^n)$, as is indicated by the final node counts for QFT unitary simulation. For GHZ, graph state and QFT unitary simulation, FTDD achieves $37\times$ speedup on average over GTN, when both approaches exploit TN optimizations.

\section{Conclusion}

This paper presents FTDD, an open-source QCS framework that incorporates a novel TN rank simplification algorithm and near-optimal contraction order search as preprocessing steps, and TDDs along with BDD optimizations for the global contraction. FTDD implements TDD operations in C++ for efficiency, and introduces edge-centric data structures to replace the cumbersome structures used in prior TDD approach, which carries redundant auxiliary variables through recursive operations. Our results demonstrate the effectiveness of FTDD, achieving an average speedup of 37$\times$ over Google's TensorNetwork library on redundancy-rich circuits, and speedups of 25$\times$ and 144$\times$ over QMDD and prior TDD approach, respectively, on Google random quantum circuits.

\bibliographystyle{IEEEtran}
\bibliography{ref.bib}

\end{document}